\documentclass[draft]{agujournal2019}
\usepackage{url} 
\usepackage{lineno} 
\usepackage[inline]{trackchanges} 
\usepackage{soul}
\usepackage{graphicx}
\draftfalse

\begin{document}

\title{Properties of Relativistic Bouncing Microbursts}
\authors{Wyatt Wetzel\affil{1}, John Sample\affil{1}, Eric Engel\affil{1}, Mykhaylo Shumko\affil{2}}

\affiliation{1}{Montana State University, Bozeman, Montana, USA}
\affiliation{2}{Johns Hopkins University Applied Physics Laboratory, Laurel, Maryland, USA}
\begin{keypoints}
\item A large collection of bouncing microbursts from SAMPEX are found
\item A distribution of scale sizes for bouncing microbursts shows that most are a few tens of kilometers in size
\item There are roughly equal numbers of one way and two way bouncing microbursts observed in this study
\end{keypoints}

\begin{abstract}
Microbursts are short duration intensifications in precipitating electron flux that are believed to be a significant contributor to electron losses in the magnetosphere.
Microbursts have been observed in the form of bouncing electron packets, which offer a unique opportunity to study the properties of microbursts and their importance as a loss process.
We present a collection of bouncing microbursts observed by the HILT instrument on SAMPEX from 1994-2004.
We analyze the locations of the bouncing microbursts in L and MLT and find they align well with the properties of relativistic microbursts as a whole.  
We find that that the majority of bouncing microbursts observed by SAMPEX have  scale sizes of $\sim 30$km at the point of observation, or $\sim 1500$km when mapped to the equator.
The time separation between the peaks of these bouncing microbursts is usually either half a bounce period or a whole bounce period.
\end{abstract}

\section{Introduction}
Radiation belt electron populations are determined by a variety of source and loss processes. 
Quantifying these processes is vital to our understanding of space weather, and to the predictive power of our modeling of the near Earth plasma environment. 
An important loss process is pitch angle scattering of electrons by plasma waves, which depletes the radiation belts by losing electrons to the atmosphere \cite{MillanReview}.
Microbursts are short duration ($<100ms$) \cite{microDuration} increases in electron flux scattered into the loss cone.
It is commonly accepted that scattering by whistler mode chorus waves is the primary cause of microbursts, based on observations  (\cite{KerstenMicroWhistler} , \cite{LorentzenMicroChorus}  \cite{BrenemanMicroWhistlerObs}).
Microbursts were first observed in the form of bremsstrahlung x-rays by balloon-borne detectors (\cite{balloon_burst_first}, \cite{andersonMicroName}).
Studies have demonstrated that microbursts are an important loss process, capable of emptying the radiation belts on the order of a day \cite{micro_loss}.
To properly constrain the losses from microbursts we must know the distribution of their scale sizes. 

Microbursts are sometimes observed in trains that descend in amplitude.
These are known as ``bouncing microbursts'' as they appear to be the same packet of electrons being observed multiple times as they bounce between their mirror points \cite{Blake96}.
They are characterized in time series data by several microbursts separated by one bounce period, each decreasing in amplitude as more electrons are lost to the atmosphere on each bounce.
The time between each microburst could also be half a bounce period, if two microbursts are generated from the same event and are launched northward and southward.  These two-way microbursts have been modeled in \citeA{bouncingModel}.

Bouncing microbursts offer a unique opportunity to measure properties of the microburst, such as  the microburst's scale size and  the amount of flux lost by the microburst to the atmosphere.
Bouncing microbursts have been observed by SAMPEX \cite{Blake96} and the FIREBIRD-II cubesats \cite{fire_bounce}, from which estimates of the scale size of the microbursts were obtained.
This study aims to build a collection of bouncing microbursts from which statistical properties of the microbursts seen by SAMPEX can be obtained.

\begin{figure}
\includegraphics[width=\textwidth]{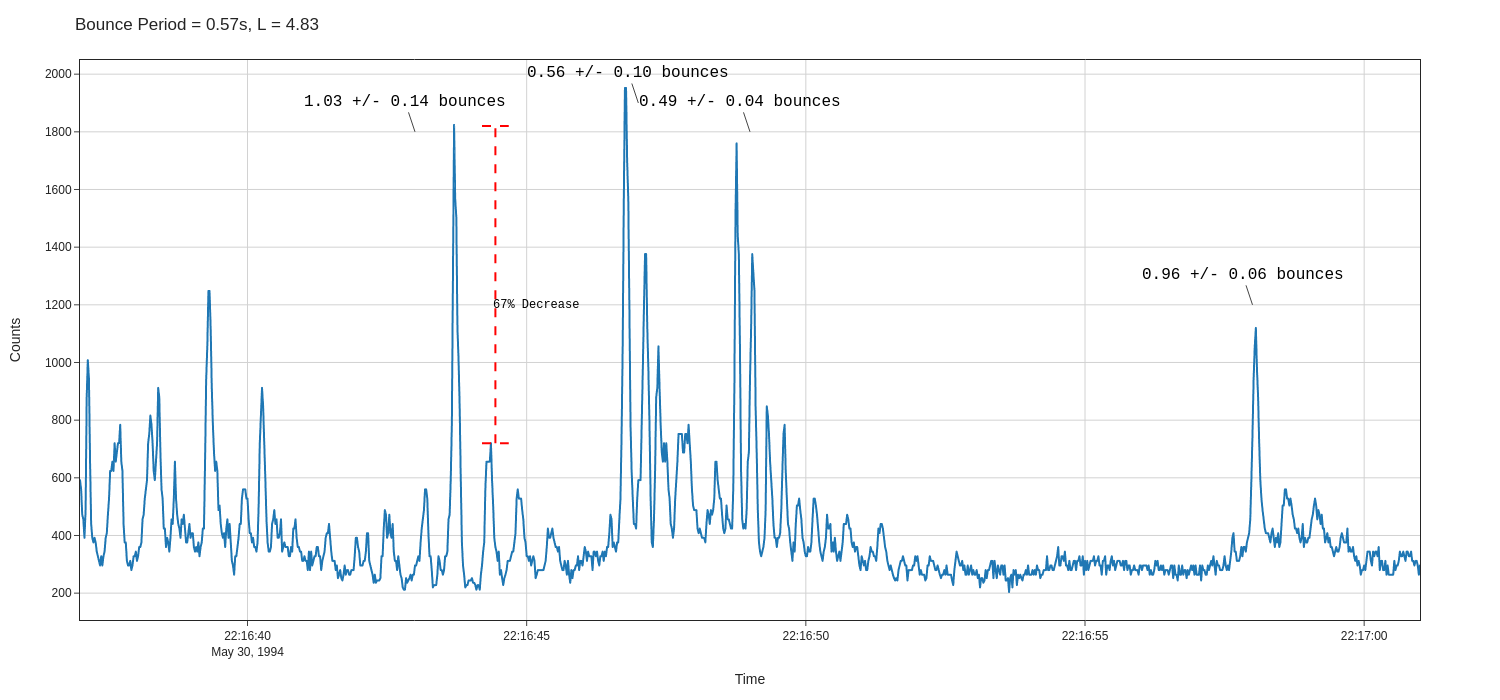}
\caption{A series of bouncing microbursts in succession. The time between peaks is labeled in units of the bounce period for $1$MeV electrons. We see four bouncing microbursts in this section of SAMPEX data. Two of them have a time separation between successive peaks of one bounce period, and two of them one half a bounce period.  }
\label{fig:case_study}
\end{figure}

\section{Bouncing Microburst Algorithm}

In this study we use the $20$ms count rate data from the  HILT instrument \cite{HILT} on the SAMPEX satellite \cite{SAMPEX}.
SAMPEX was in a high inclination LEO for the duration of its mission.
The HILT instrument consisted of sixteen SSD's and was designed to detect heavy ions, but was also sensitive to electrons with energies greater than $1$MeV.

To search for bouncing microbursts, we construct an algorithm based on the cross correlation of SAMPEX data with a sample ''kernel`` that is shaped like a typical bouncing microburst.
These kernels are constructed of Gaussians with various separations ranging from $.05$s to $1$s to attempt to capture bouncing microbursts with a variety of bounce periods.
We then take the cross correlation of the SAMPEX time series data and each of the kernels, and flag times with high correlation for manual review.
We also require that each flagged location contain a microburst according to the parameter described in \citeA{obrien_parameter}.

Our algorithm would often identify clusters of microbursts that did not meet our criteria to be considered microbursts, such as groups with uneven spacing.
This led to a high false positive rate which necessitated reviewing the data the algorithm identified manually.
The algorithm initially identified 7107 candidates, which were narrowed down to 1355 bouncing microbursts by hand.

Several examples of bouncing microbursts in succession are shown in Fig. \ref{fig:case_study}.
There are four bouncing microbursts identifiable in this section of data; two of these are characterized by peak to peak separations of one bounce period and two by half a bounce period.

\section{Characteristics}
With this collection of bouncing microbursts, the statistical properties of these events can be studied. 
These events occur primarily on the morningside from $L=4$ to $L=6$, which matches the characteristics of microbursts observed by SAMPEX as a whole \cite{burst_occurrence}.

\begin{figure}[h!]
\includegraphics[width=\textwidth]{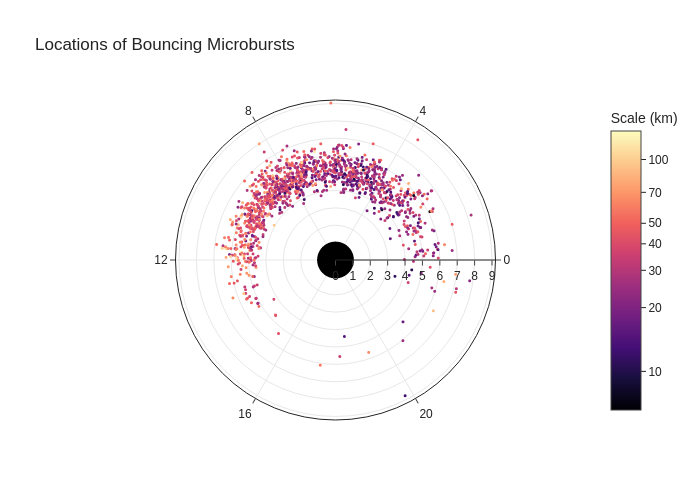}
\caption{Locations of bouncing microbursts in L and MLT. Color scale indicates local azimuthal scale size.  }
\label{fig:dial}
\end{figure}

\begin{figure}[h]
 \includegraphics[width=\textwidth]{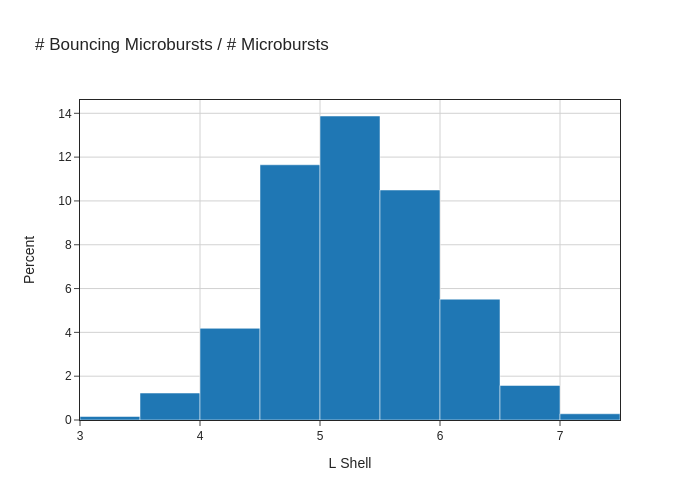}
 \caption{Comparison of bouncing microburst observations to total number of microbursts as defined by the O'Brien parameter.  }
 \label{fig:fraction}
\end{figure}

The bouncing microbursts observed in this study are a fraction of the total number of microbursts observed by SAMPEX.
A comparison between the total population of microbursts and the subset of microbursts in this paper is shown in Fig. \ref{fig:fraction}.
Microbursts from the total set of SAMPEX data are found using the O'Brien parameter, then compared to the number of bouncing microbursts in each L-shell.

\begin{figure}[h]
 \includegraphics[width=\textwidth]{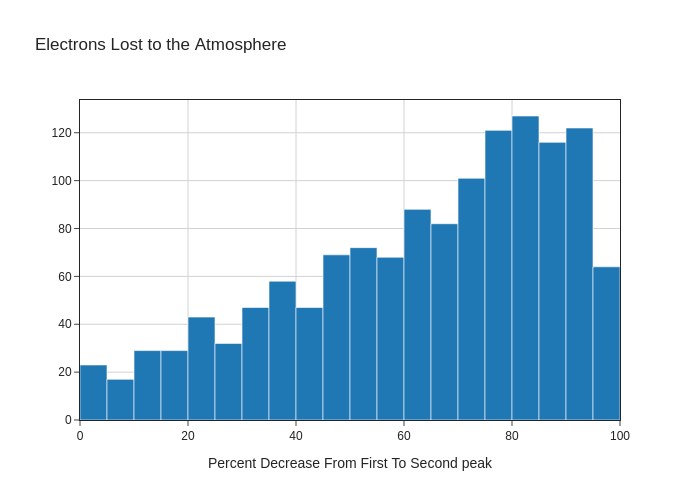}
 \caption{Histogram of percent difference in counts between first and second peak in a bouncing microburst.}
\label{fig:percentDecrease}
\end{figure}

With this collection, we can estimate the fraction of electrons lost to the atmosphere and the amount backscattered on each bounce. \citeA{ThorneTimescale} estimated the fraction of electrons surviving each bounce to be $.675$ (or a decrease of $32.5\%$) based on an observation of a bouncing microburst in \citeA{Blake96}.
We take the percent difference in count rate from the first peak to the second peak in a bouncing microburst and plot the resulting distribution in Fig. \ref{fig:percentDecrease}.
The flux lost on the first bounce is widely distributed but more heavily weighted towards a loss of  around $80\%$.
Simulations \cite{Backscatter} have found substantial energy lost by electrons backscattered from the atmosphere, so some electrons with enough energy to be observed by HILT initially could have dropped below the $1$MeV detection threshold, thus biasing Fig. \ref{fig:percentDecrease} towards higher percentages of electrons lost.

\section{Bounce Period Distribution}

\begin{figure}[h!]
\includegraphics[width=\textwidth]{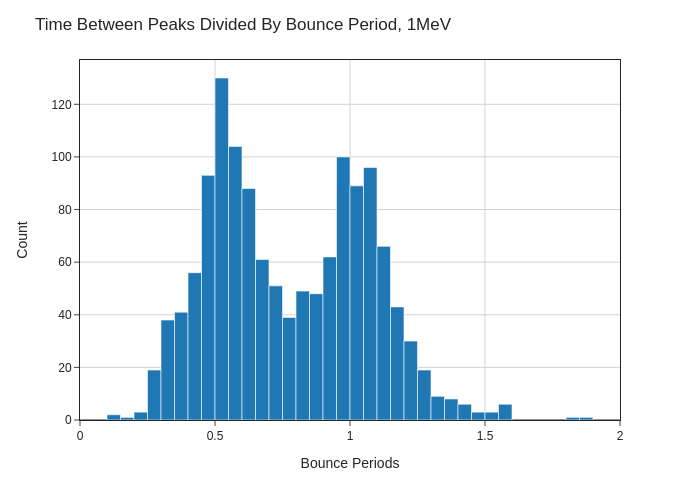}
\caption{Distribution of time between peaks measured in bounce periods.}
\label{fig:periods}

\end{figure}

To calculate the bounce period of the electrons that make up the bouncing microbursts, we assume that all electrons have an energy of $1$MeV, are at the edge of the loss cone, and mirror at $100$km altitude.
We calculate the L shell of the microbursts by using SAMPEX's reported location and the Tsyganenko-Sitnov (TS05) model \cite{TS05} to find the bounce period for each microburst (Fig. \ref{fig:periods}).
We find that the time separation between successive peaks in a bouncing microburst tends to be about one bounce period or one half of a bounce period.

\section{Scale Sizes}
Measurements of bouncing microbursts allow us to estimate the local azimuthal scale size of the microburst, and from that the equatorial scale size.
The local scale size of each microburst (in units of Earth's radius) is computed as

\begin{equation}
 l_{obs} = T v_D  = T \frac{2 \pi}{\tau_D} (1+ \frac{A}{R_e}) \cos \lambda 
\end{equation}
where $T$ is the time between the first and last bounce observed, $v_D$ is the electron drift velocity, $\lambda$ is the magnetic latitude, $A$ is the altitude of SAMPEX at the observation, $R_e$ is the radius of the Earth, and $\tau_D= (60)(62.7)(\frac{MeV}{E})(\frac{1}{L})s$ is the drift period of an electron at an energy $E$ and L-shell $L$ \cite{ParksTextBook}.
\begin{figure}[h!]
 \includegraphics[width=\textwidth]{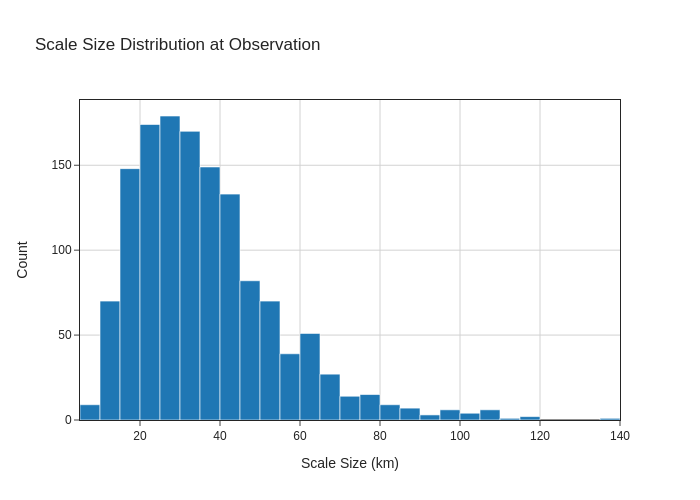}
 \caption{Local scale size, calculated at SAMPEX's location at time of observation.}
  \label{fig:local}

\end{figure}
The resulting distribution of local scale sizes of the microbursts we have collected is presented in Fig. \ref{fig:local}.

We can convert the scale size at the location of the observation to the microburst's size at the equator with Gauss' law, setting the flux at SAMPEX's location equal to the flux at the equator:  $B_{eq} l_{eq}^2 = B_{obs} l_{obs}^2$, where $B_{eq}$ and $B_{obs}$ are the magnitudes of the magnetic field at the equator and at SAMPEX, and $l_{eq}$ and $l_{obs}$ are the scale sizes of the microburst at the equator and at SAMPEX.

Idealizing Earth's magnetic field as a dipole, we find that the equatorial scale size of the microbursts is 

\begin{equation}
l_{eq} = l_{obs} \sqrt{2} \left( \frac{L R_e}{r_{obs} } \right)^{3/2}
\end{equation}

where $r_{obs}$ is the distance from the center of the Earth to the location of the observation.

\begin{figure}[h!]`
\includegraphics[width=.9\textwidth]{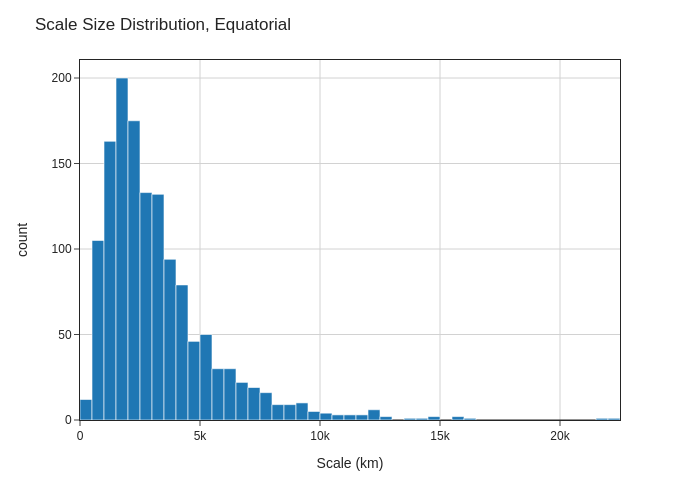}
\caption{Histogram of equatorial azimuthal scale sizes.}
\label{fig:sep_hist_eq}
\end{figure}

Bouncing microburst scale sizes tend to be on the order of a few tens of kilometers, or around  $1000$km when mapped back to the magnetic equator. \citeA{fire_bounce} observed a bouncing microburst with the FIREBIRD cubesats with a scale size of $51$km, within the distribution found in this paper.

With these distributions of scale size, we can estimate the typical flux lost to the atmosphere  by a bouncing microburst.
These microbursts typically correspond to $O(10^3)$ counts; using SAMPEX's geometric factor of $60$ cm\textsuperscript{2} sr and a local scale size of about $30$km we find $O(10^{15})$ electrons are lost per typical bouncing microburst.
For comparison, differential electron flux data from RBSPICE is used to estimate the number of electrons above $1MeV$ passing through a  $500$km wide strip at the magnetic equator.
We find $O(10^{19})$ electrons in this strip, meaning a typical bouncing microburst from this collection would empty about $.01 \%$ of the MeV electron population from that L-shell.

\section{Summary and Conclusions}
Using an algorithm based on taking the cross correlation of years of SAMPEX HILT data with a bouncing microburst shaped kernel we have identified 1355 bouncing microbursts.
These bouncing microbursts are distributed similarly to the larger set of microbursts observed by SAMPEX \cite{burst_occurrence}.
We find that the amount of electrons lost from the first peak in a bounce to the second is widely distributed, weigthed towards $80\%$ of counts lost from peak to peak.

The time separation between observed peaks of a bouncing microburst is distributed roughly equally around a half bounce period and a whole bounce period.
There are therefore similar numbers of ``one way'' and ``two way'' microbursts observed in this study, i.e. microbursts generated singly or as a pair northward and southward traveling packets.

Additionally, from this collection we find that the scale sizes of these bouncing microbursts are usually between $10$km and $60$km.
A previous study using the FIREBIRD cubesats  found a bouncing microburst with local scale size of $51$km \cite{fire_bounce}, well within the distribution determined in this study.
\cite{Blake96} observed that the relative infrequency of bouncing microbursts compared to solitary bursts indicates that most microbursts must have scale sizes of less than a few tens of kilometers, which was the minimum size of microburst that SAMPEX could observe bouncing.
The bouncing microbursts in this paper therefore represent the population of larger microbursts, with smaller microbursts being difficult to observe bouncing with SAMPEX.
The bouncing microbursts in this study were roughly $10\%$ of the total number of microbursts observed by SAMPEX around $L=5$.
These scale sizes are similar to the scale sizes of whistler mode chorus waves \cite{chorus_scales}, the primary candidate for the waves that generate microbursts.

\section*{Open Research Section}
HILT data products and SAMPEX orbit data are available from \url{http://www.srl.caltech.edu/sampex/DataCenter/data.html}.

\acknowledgments
This work was supported by NASA grant 80NSSC19K0702, and by NASA grant 80NSSC18K1275.

\bibliography{master}

\end{document}